\documentclass[11pt, bibliography=totocnumbered,headsepline,abstract=true]{scrartcl}
\usepackage[paper=a4paper, left=25mm, right=25mm, top=25mm, bottom=30mm,marginparwidth=2.75cm]{geometry}

\usepackage{amsmath, amssymb, graphicx, url,hyperref}

\usepackage[utf8]{inputenc}
\usepackage{soul}
\usepackage{xcolor}
\usepackage{enumitem}   
\usepackage[backend=bibtex, style=science, citestyle=numeric-comp, sorting=none, maxnames=3, sortcites=true, date=year, isbn=false, doi=true, url=false]{biblatex}
\newcommand\blankfootnote[1]{%
	\begingroup
	\renewcommand\thefootnote{}\footnotetext{#1}%
	\addtocounter{footnote}{-1}%
	\endgroup
}

\begin{document}
	
	\title{Probing the Disorder inside the Cubic Unit Cell of Halide Perovskites from First-Principles}
	
	\author{Xiangzhou Zhu,
		Sebasti\'{a}n Caicedo-D\'{a}vila,
		Christian Gehrmann,
		\\ and David A. Egger$^*$}
	\date{} 
	\maketitle
	
	\begin{center}
		Department of Physics, Technical University of Munich, Garching, Germany
	\end{center}
	\blankfootnote{$^*$Email Address: david.egger@tum.de}
	
	\begin{abstract}
		\small
		Strong deviations in the finite temperature atomic structure of halide perovskites from their average geometry can have profound impacts on optoelectronic and other device-relevant properties.
		Detailed mechanistic understandings of these structural fluctuations and their consequences remain, however, limited by the experimental and theoretical challenges involved in characterizing strongly anharmonic vibrational characteristics and their impact on other properties.
		We overcome some of these challenges by a theoretical characterization of the vibrational interactions that occur among the atoms in the prototypical cubic CsPbBr$_3$.
		Our investigation based on first-principles molecular dynamics calculations finds that the motions of neighboring Cs-Br atoms interlock, which appears as the most likely Cs-Br distance being significantly shorter than what is inferred from an ideal cubic structure. 
		This form of dynamic Cs-Br coupling coincides with very shallow dynamic potential wells for Br motions that occur across a locally and dynamically disordered energy landscape. 
		We reveal an interesting dynamic coupling mechanism among the atoms within the nominal unit cell of cubic CsPbBr$_3$ and quantify the important local structural fluctuations on an atomic scale.
	\end{abstract}
	
	\pagebreak

\section*{Introduction}
Halide perovskites (HaPs) are crystalline semiconductors with very useful optoelectronic properties for technological applications in photovoltaics and light-emitting devices\cite{snaith2013,green2014,brenner2016,stranks2015,correa-baena2017,stoumpos2016,nayak2019,li2017}.
At the same time, these systems exhibit lattice vibrational characteristics that are unusual when compared to canonical inorganic semiconductors (\textit{e.g.}, Si or GaAs), which motivates a variety of interesting scientific questions for the research community.\cite{egger2018}
A key observation in this context is the dynamic symmetry breaking that occurs around room temperature in these materials:\cite{beecher2016,yaffe2017}
while the \textit{average structure} of many HaPs with ABX$_3$ stoichiometry exhibits a high symmetry at around room temperature and above,\cite{moller1958,weber1978,saparov2016} various studies have found that locally and instantaneously the \textit{actual atomic geometry} strongly deviates from the idealized, averaged one.\cite{worhatch2008,beecher2016,bernasconi2017,page2016,yaffe2017}
The structural fluctuations can involve both \textit{A}- and \textit{X}-site ions as expressed in, \textit{e.g.}, octahedral distortions that were found to be concomitant with significant Cs motion,\cite{straus2020} 2D overdamped fluctuations of halides\cite{lanigan-atkins2021}, an X-Pb-X scissoring mode,\cite{bird2021} \textit{A}-site \textit{X}-site dynamic coupling,\cite{singh2020,ghosh2017} as well as large transversal displacements of the halides.\cite{gehrmann2021}
Concurrent with these local, dynamic fluctuations in the structure are octahedral tilting instabilities occurring in various relevant candidate materials of this class.\cite{gao2021,yang2017,klarbring2019,menahem2021}
Therefore, the appearance of a finite-temperature atomic structure that is locally and dynamically disordered, while being long-ranged ordered at the same time, represents a generic feature of a wide range of HaP materials.\\
The local fluctuations in the atomic structure are important because they impact the optoelectronic properties of HaPs at finite temperature and, by extension, the characteristics of devices made from them. 
Perhaps most relevant is the case of finite-temperature charge-transport properties of HaPs, since it has been shown that the unusual lattice vibrations go hand-in-hand with a dynamic disorder and transient localization of band-edge carriers \cite{mayers2018,lacroix2020,schilcher2021,wright2021}.
Closely related to this are the intriguing interconnections between these structural fluctuations and signatures of polaronic and light-induced effects in HaPs that are currently discussed in the literature.\cite{wu2017,lanigan-atkins2021,guzelturk2021,schilcher2021,limmer2020,buizza2021,iaru2021,thouin2019,feldmann2019}
Moreover, the dynamic structural fluctuations occurring in HaPs were shown to influence defect characteristics of these materials as well and might bear important implications for defect tolerance and self-healing effects.\cite{cohen2019,wang2019,li2019,kumar2020,zhang2020,chu2020,cahen2021,nan2018}\\
Therefore, obtaining a microscopic picture of the involved mechanisms is key to an improved understanding of the physical properties of HaPs, but it also presents various challenges for theoretical and experimental studies alike.
One reason is that direct detection of the consequences of dynamic symmetry breaking requires probing observables related more to the instantaneous structure, rather than the average one, with techniques that are sensitive in both the time and spatial domain.
From a theory perspective, it has been difficult obtaining refined understandings of finite-temperature dynamics of HaPs and potentially important effects involved in it, such as temperature-dependent vibrational coupling.
This is in part because such would require leaving aside a harmonic phonon picture in which vibrational modes are treated entirely uncoupled \cite{ziman2001}.
Overcoming these challenges will support the identification of the mechanisms that are active in HaPs and impact their functional properties at finite temperature.\\
In this article, we take a step in the direction of resolving the mechanisms that impact the finite-temperature vibrational dynamics of HaPs.
Our focus is on the vibrational interactions that occur locally among the atoms in the prototypical cubic CsPbBr$_3$ at a temperature of $425~\mathrm{K}$ and $525~\mathrm{K}$, using first-principles molecular dynamics (MD) calculations based on density-functional theory (DFT),  {which capture the local instantaneous structures as they occur at finite temperature when measured experimentally in, \textit{e.g.}, Raman or infrared spectroscopy.}
 {The cubic phase of CsPbBr$_3$ is considered because it shows vibrational effects that are especially interesting, such as local polar fluctuations,{\cite{yaffe2017}} and hence represents a greater number of compounds showing structural fluctuations as have been discussed above.
Our approach is to investigate the dynamics of atom types and their interconnections to those of other atom types, rather than considering specific bonds present in the system.
With this, we find a surprising} interlocking in the dynamics of neighboring Cs and Br atoms, which manifests in the most likely Cs-Br distance occurring in the system at  {temperatures of} $425~\mathrm{K}$  {and $525~\mathrm{K}$} to be \textit{significantly shorter than what is inferred from an ideal cubic structure}. 
This form of dynamic Cs-Br coupling occurring inside the nominal unit cell of CsPbBr$_3$ is found to be related to the in-phase octahedral tilting dynamics and to coincide with very shallow potential wells for Br displacements that we obtain directly from MD data. 
The fact that the Br motions are entangled with those of neighboring Cs atoms is shown to be connected to a locally and dynamically disordered energy landscape for Br atoms, which is formed by a multitude of potential wells that are close in energy.
By probing the disorder inside CsPbBr$_3$ and revealing a coupling in the Cs-Br dynamics, our study quantifies the profound deviations of the local structure from the ideal and averaged one on an atomic scale.

\section*{Results and Discussion}

\begin{figure*}
    \centering
    \includegraphics[width=.8\linewidth]{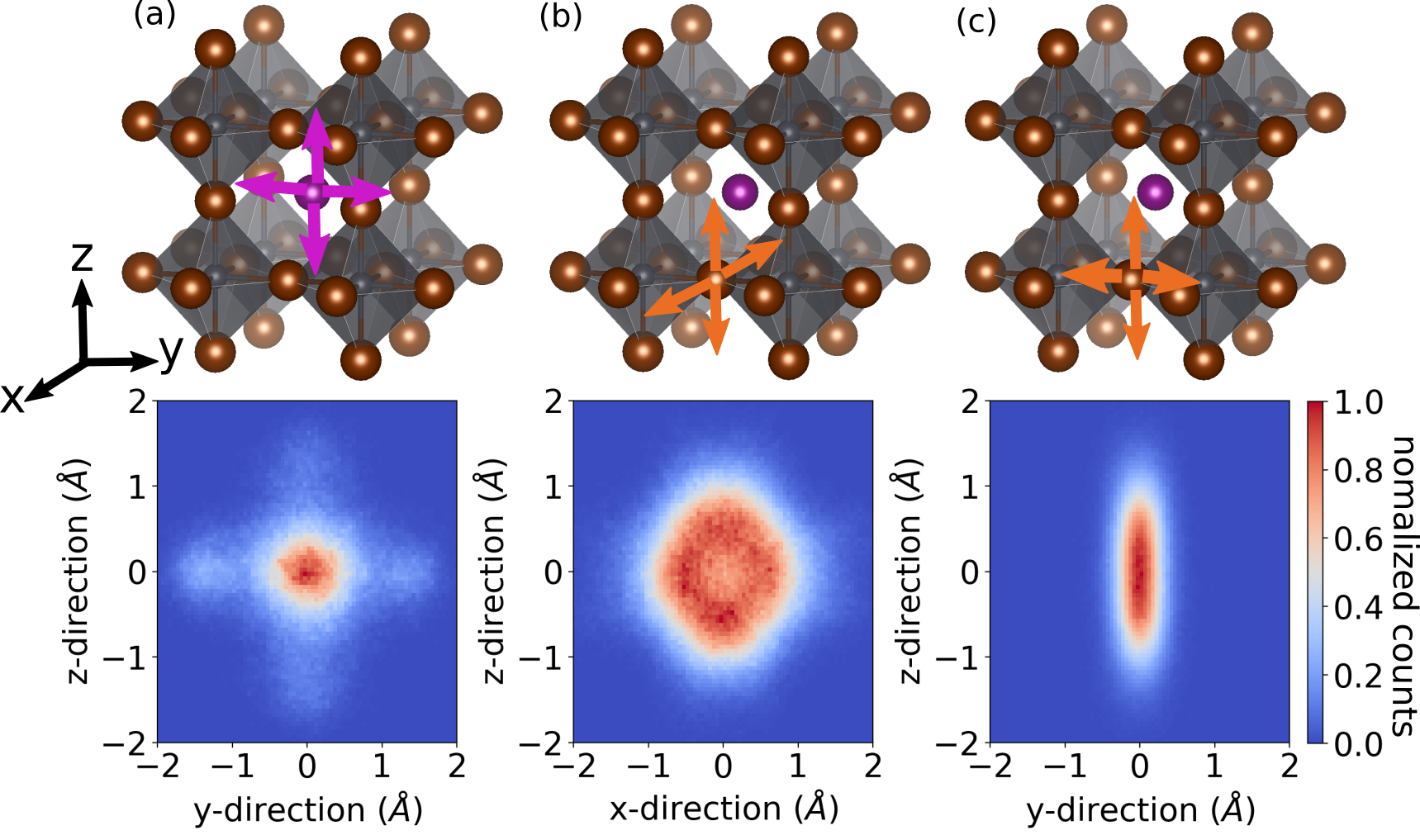}
    \caption{
    2D histograms of Cs displacements, occurring  {in the MD calculations at $T=425~\mathrm{K}$}, on the $yz$-plane (panel \textbf{(a)}), Br displacements on the $xz$-plane (panel \textbf{(b)}), and Br displacements on the $yz$-plane (panel \textbf{(c)}).
    Top panels are schematic illustrations showing the structure of CsPbBr$_3$ and the directions chosen for the 2D histograms that are shown in the bottom panels. Note that the origin of the 2D histograms corresponds to the ideal lattice position of the considered atom.
    }
    \label{fig:csbr_dist}
\end{figure*}

\begin{figure}[t]
	\centering
	\includegraphics[width=0.5\linewidth]{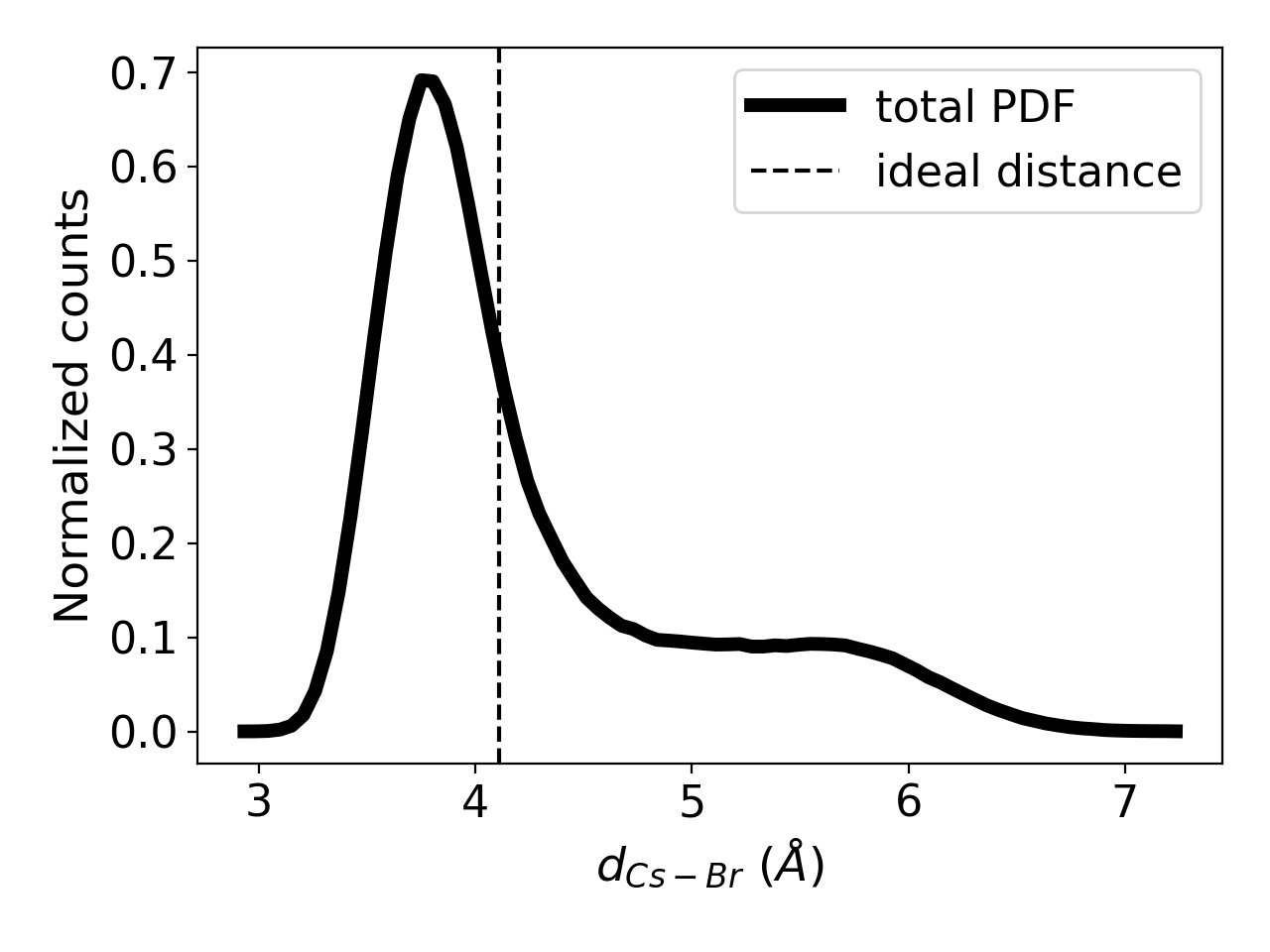}
	\caption{
		Pair-distribution function (PDF) of nominal nearest-neighbor Cs and Br atoms, calculated from MD at $T=425~\mathrm{K}$ (see text for details) as a function of the Cs-Br distance, $d_\mathrm{Cs-Br}$.
		The vertical, dashed line marks $d_\mathrm{Cs-Br}$ that can be inferred using the ideal, cubic unit cell of CsPbBr$_3$.
		{See SI for the Cs-Br PDF computed from MD at $T=525~\mathrm{K}$.}
	}
	\label{fig:pdf}
\end{figure}

In our work, we  {apply MD calculations based on DFT (see Methods section for further details), which allows us to} focus on the local atomic structure as it occurs in CsPbBr$_3$ at $425~\mathrm{K}$, and first investigate the Br and Cs displacements separately.
Using the ideal cubic structure as a reference, occurrences of Br and Cs displacements are recorded in 2D histograms across the planes illustrated in the top panels of Fig.~\ref{fig:csbr_dist}a-c.
If the vibrational dynamics in the system would occur in a fully harmonic potential energy landscape, the displacement distribution functions would be Gaussians peaked at a zero displacement marking the ideal lattice position.\\
The distribution for the Cs displacements (see Fig.~\ref{fig:csbr_dist}a, bottom panel) is indeed peaked at a zero displacement that corresponds to the center of a nominally cubic cell. 
However, the Cs distribution also shows six side peaks (see also  {Fig. S1 in Supporting Information, SI, for 1D histograms}) that correspond to Cs motions towards the face centers of such a cubic cell, which means that it certainly does not correspond to a Gaussian function; note that only four of the six peaks are shown. 
We also note that the Cs motion to the six side peaks can be associated with the head-to-head motion of Cs discussed in Ref. \cite{yaffe2017} (see SI, Fig. S4).
The distribution of the generally very large Br displacements occurring perpendicular to the Pb-Br-Pb bonding axis (see Fig.~\ref{fig:csbr_dist}b, bottom panel) is found to not be Gaussian as well and also to peak at a nonzero displacement (see Fig. S2 in SI for additional analysis).
These features in the Cs and Br displacement distributions are yet another manifestation of the known effect of anharmonicity in the vibrations of this and other HaP systems. \cite{liu2019,lahnsteiner2021,yang2017,lanigan-atkins2021,klarbring2019,gao2021,marronnier2017,klarbring2018}
We can further support this rationale by examining Br displacements occurring in parallel to the Pb-Br-Pb bonding axis (the $y$-direction in Fig.~\ref{fig:csbr_dist}c, bottom panel), which are known to be involved in vibrations that are far less anharmonic:\cite{lanigan-atkins2021,gehrmann2021} 
the distribution of these Br displacements shows that they are much smaller in magnitude, peak at approximately a zero displacement, and follow a Gaussian function (see Fig. S3 in SI).\\
The fact that Cs displacements are peaked at zero while Br displacements are peaked significantly away from zero, implies that the most likely nearest-neighbor distance between the two species is \textit{not the one found in the ideal cubic structure}.
To test and further explore this hypothesis, we consider Cs and Br together and first compute the pair distribution function (PDF) \cite{billinge2019} for all possible first nearest-neighbor pairs of Cs and Br atoms, a quantity that has been discussed for HaPs and their structural fluctuations.\cite{zhao2020} 
The Cs-Br PDF yields the probability of finding a nearest-neighbor Br atom at a distance $d_\text{Cs-Br}$ from a Cs atom and \textit{vice versa}.
Because our goal is to analyze the disorder within the nominal unit cell, we count the 12 nominal first-nearest neighbor Br atoms of each Cs (see Methods section for further details).
From Fig.~\ref{fig:pdf} we find that the peak in the Cs-Br PDF deviates significantly from the nearest-neighbor distance that is expected for the ideal cubic structure, indicating a most likely  {$d_\text{Cs-Br}\approx3.8$~\AA} ~compared to  {$d_\text{Cs-Br}\approx4.1$~\AA} ~for the ideal cubic case, which we obtain from the unit cell relaxed in a static DFT calculation. 
Interestingly, the PDF also shows a pronounced tail ranging up to $d_\text{Cs-Br}\approx 6.5$~\AA.\\
 {The finding that \textit{actual Br-Cs distances} can be very different compared to the one that is inferred from an ideal cubic structure by itself is perhaps not entirely surprising.
However, the results of Figs.~{\ref{fig:csbr_dist}} and {\ref{fig:pdf}} establish more than that:}
because the ideal cubic structure is an average geometry in both the time and spatial domain, for Cs-Br pairs that are sufficiently far apart from one another the Cs-Br PDF is expected to show peaks at multiples of $d_\text{Cs-Br}$ of the ideal structure.
And yet, we find that the nearest-neighbor Cs-Br pairs in most cases appear at distances that are significantly shorter than what is observed in an ideal cubic structure. 
Therefore, these results suggest an interesting coupling mechanism active in the joint dynamics of Br and Cs atoms. 
For comparative purposes, we show the Cs-Br PDF at $525~\mathrm{K}$ (Fig. S5) as well as the Pb-Br PDF at $425~\mathrm{K}$ and $525~\mathrm{K}$ (Fig. S6) in the SI.\\
\begin{figure*}
	\centering
	\includegraphics[width=.9\textwidth]{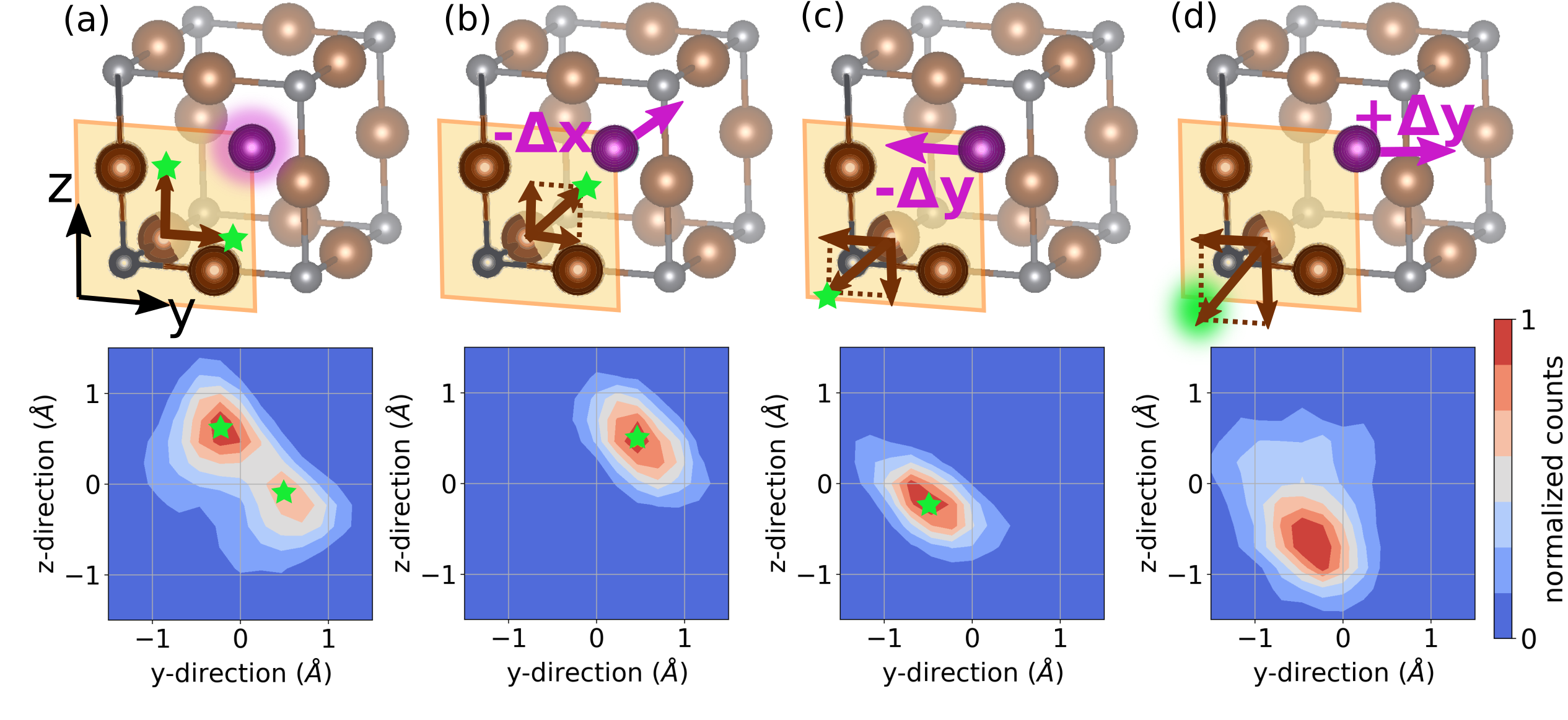}
	\caption{
		Conditional 2D histograms of Br displacements occurring under the condition that a nearest-neighbor Cs atom exhibits a specific displacement type during the MD at $T=425~\mathrm{K}$, which are organized along panel \textbf{(a)}-\textbf{(d)} (see text for details). Top panels are schematic illustrations of the Cs displacement type, shown in violet, and the preferred motion of a nearest-neighbor Br atom occurring while that specific Cs displacement is active, shown in brown. The histograms of Br displacements shown in the lower panels are tracked on the planes indicated in the upper panels. Green symbols refer to the most frequently occurring region of a nearest-neighbor Br displacement in each case.
	}
	\label{fig:Dyn5d}
\end{figure*}
To probe the coupled motion of Cs and Br atoms, we focus on nearest-neighbor Cs-Br pairs and compute 2D conditional histograms: 
they record the distribution of Br displacements occurring across specific planes under the condition that a nearest-neighbor Cs atom moves in a specific way.
Hence, with this we can assess which type of Cs and Br motions are present in the crystal at the same time and how they influence one another.
For example, for independent nearest-neighbor Cs-Br motion one would expect very similar Br displacement distributions irrespective of the chosen type of Cs displacement setting the condition.
We consider a set of conditions for Cs that are motivated by our above result (see Fig.~\ref{fig:csbr_dist}a) that Cs either oscillates around the center of a nominal cubic cell or around 6 preferred positions due to a large displacement that is occurring along either one of the $\pm x, \pm y, \pm z$ directions.
All conditioning cases following from this can be mapped by symmetry (see Fig. S7 in SI) to consideration of the following 4 scenarios of Cs displacement types acting as conditions when recording Br displacement histograms.
 {The 4 scenarios are illustrated in the top panels of Fig.~{\ref{fig:Dyn5d}}, described in detail in the Methods section and briefly summarized here:}
In case (i), Cs oscillates around a cubic center (see Fig.~\ref{fig:Dyn5d}a).
Case (ii) stipulates that Cs undergoes a large displacement parallel to the Pb-Br-Pb bonding axis (the ($+x$, $-x$) direction in Fig.~\ref{fig:Dyn5d}b).
For case (iii), Cs is considered to exhibit a large displacement on the plane perpendicular to the Pb-Br-Pb bonding axis towards one nominal nearest-neighbor Br position of the ideal cubic cell (the ($-y$, $-z$) direction in Fig.~\ref{fig:Dyn5d}c).
Finally, case (iv) is the same as (iii), but  {the difference is that} Cs moves away from the nominal Br position in the ideal cubic cell (the ($+y$, $+z$) direction in Fig.~\ref{fig:Dyn5d}d).
We note that scenarios (ii)-(iv) involve a large displacement of Cs according to the side-peaks of Fig.~\ref{fig:csbr_dist}a.\\
In scenario (i), when Cs oscillates around the center of a nominally cubic cell (see Fig.~\ref{fig:Dyn5d}a), the Br histogram shows a broad distribution with two peaks that indicate two preferred Br positions around ($y\approx0.1$~\AA, $z\approx0.5$~\AA) and ($y\approx0.5$~\AA, $z\approx0.1$~\AA), which can be shown to correspond mainly to an octahedral rotation around one axis (see Fig. S8 in SI).
In scenario (ii), when Cs is displaced in-parallel to the Pb-Br-Pb bonding axis (see Fig.\ref{fig:Dyn5d}b), in contrast to case (i) the Br histogram displays only a single peak (see Fig.~\ref{fig:Dyn5d}b). 
The peak signifies a Br displacement over the $yz$-plane, which corresponds to octahedral rotation around both the $y$- and the $z$-axes (see Fig. S8 in SI).
The result of scenario (iii), which probes Br displacements when Cs moves along a direction perpendicular to the Pb-Br-Pb axis towards a Br lattice position (see Fig.~\ref{fig:Dyn5d}c), finds the histogram for Br displacements changing to the opposite quadrant on the $yz$-plane.
Last is scenario (iv), where Cs moves along a direction perpendicular to the Pb-Br-Pb axis, but  {in contrast to case (iii) it now moves} away from a Br lattice position, in which case we find the Br histogram to entail substantially larger displacements, a wider distribution, and an absence of a clear peak position.\\
Taken together, the findings clearly show that Cs and Br displacements are interlocked, because both the shape and peaks of the Br displacement distribution strongly depend on the type of nearest-neighbor Cs displacement active in the system.
 {This is unexpected because A-site cations are often assumed to not interact with the rest of the HaP lattice, which we find is not the case because statistical distributions for Cs and Br displacements are not independent from one another.}\\
\begin{figure}[t]
 	\centering
 	\includegraphics[width=0.6\linewidth]{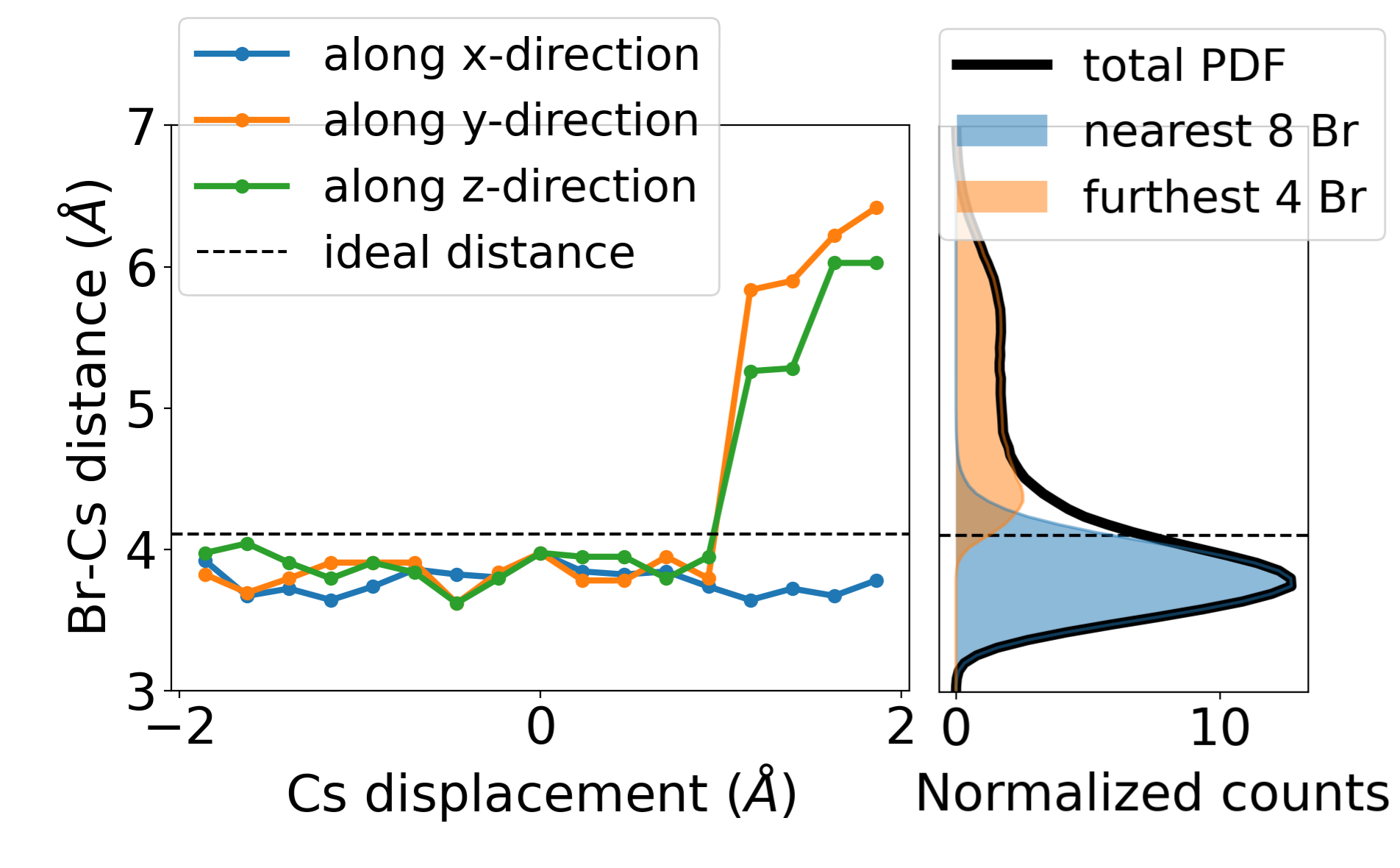}
 	\caption{
 		\textit{Left.} The most frequent Cs-Br distance, occurring in MD at 425 K, calculated as a function of the Cs displacement along the three crystalline directions, such that a value of $d_\text{Cs-Br}=0$ corresponds to Cs located at the nominal lattice position. The $x$-direction (blue curve) is parallel to the Pb-Br-Pb bonding axis whereas the $y$- and $z$-direction (orange and green curves) span the plane perpendicular to it. Thus, along $y$- and $z$-direction, negative values represent the Cs moving towards the nominal lattice position of a nearest-neighbor Br atom (see Fig.~\ref{fig:Dyn5d}c and text), while positive values correspond to Cs moving away from it (Fig.~\ref{fig:Dyn5d}d).
 		\textit{Right.} PDF of the 12 nominal Cs-Br nearest-neighbors (black curve) and the deconvoluted components, corresponding to 8 nearest-neighboring Br atoms (blue curve) and the remaining 4 (orange curve).
 	}
 	\label{fig:fig4}
\end{figure}
Therefore, it is interesting to investigate the coupled Cs-Br motion inside the nominal unit cell of CsPbBr$_3$ further.
We extend our analysis from above, now considering the same single Cs-Br pair as in Fig.~\ref{fig:Dyn5d} but sampling a  {wider range of Cs positions in the 3D lattice when computing conditional histograms of Br displacements, which allows for a higher resolution of the Cs-Br coupling}. 
From the result of this,  {we} obtain the peak corresponding to the Br displacement with the largest number of counts for each Cs displacement. 
From it we can calculate the most frequent Cs-Br distance, occurring in the MD at $425~\text{K}$, as a function of the Cs displacement along the three crystalline directions (see Fig.~\ref{fig:fig4}, left panel).
Remarkably, the preferred Cs-Br distance is similar (around 3.8~\AA) for a considerable fraction of all considered cases.
This is in line with the finding discussed in regard to the Cs-Br PDF (\textit{cf.} Fig.~\ref{fig:pdf}), namely that the most frequent Cs-Br distance in the actual structure is significantly shorter than the Cs-Br distance inferred from the ideal cubic one.
The situation is entirely different when the Cs displacements reaches relatively large positive values on the $yz$-plane, that is when it exhibits a displacement driving it \textit{away from the Pb-Br-Pb bond axis}.
Then, the most frequent distance changes to a much larger value  ($\approx$5~\AA~to 6~\AA).
\\
Importantly, the combination of these two effects explains the main features in the Cs-Br PDF we discussed above (\textit{cf.} Fig.~\ref{fig:pdf}): our analysis based on symmetry (see Fig. S9 in SI) finds that 8 out of the 12 nearest-neighbor Br atoms of Cs must correspond to the short Cs-Br distance and the remaining 4 nearest neighbors to the much longer distance.
Therefore, we calculate the Cs-Br PDF separately for the 8 nearest-neighbor Br atoms with shorter (blue curve in Fig.~\ref{fig:fig4}, right panel) and the remaining 4 nearest-neighbor Br atoms with longer distance (orange curve Fig.~\ref{fig:fig4}, right panel).
 {The result of this procedure} shows that the former explains the peak and the latter explains the pronounced tail in the Cs-Br PDF, confirming the validity of our analysis based on conditional histograms.\\
Having established the presence of a local coupling in the Cs-Br dynamics in CsPbBr$_3$, it is tempting to ask about connections of it to other properties of the material system.
The question is addressed here by first investigating the energetics of Br motions in the system, which is important because the Br atoms are involved in the octahedral dynamics  at finite temperature and their atomic orbitals participate in the electronic states close to the band edges.
To this end, we consider the Br atoms to move in an ensemble-averaged potential that is due to the interactions with the other atoms in the system.
 {Hence,} we perform a Boltzmann inversion  {(see Methods section)} of Br displacements obtained from MD  {at $T=425$~K and $T=525$~K, in order} to compute effective potential wells for Br displacements (see Fig.~\ref{fig:invb_atom}).
 {It is insightful to compare potential wells for Br displacements at the two different temperatures in the MD, because the $425$~K MD calculations are relatively close to the known $\approx 400$~K tetragonal-cubic phase-transition of CsPbBr$_3$.}\cite{hirotsu1974}
For the sake of clarity, we focus on a specific direction in the Br displacements to obtain a 1D potential by taking an average of the $y$- and $z$-directions, \textit{i.e.}, the two directions perpendicular to the Pb-Br-Pb bonding axis (\textit{cf.} Fig.\ref{fig:csbr_dist}, top panel). 
 {It is noted that this potential well corresponds to displacements of a single Br atom embedded in the ensemble-averaged potential that is generated by all vibrational excitations present in the system at specific temperature.}
We refer to it as a "dynamic potential well", because it includes the effect of the populated vibrational modes and their anharmoncity on the Br motion.
The procedure can be compared to results from a frozen-phonon approach  {(see Methods section)}, \textit{i.e.}, energy changes computed from just the displacements involved in a particular phonon mode, which we refer to as a "static potential well" because it does not account for vibrational coupling and anharmoncity.\\
First of all, the dynamic potential wells  {at both temperatures are} found to be very shallow (see Fig.~\ref{fig:invb_atom}), \textit{i.e.}, Br displacements in regions of $\pm2~\mathrm{\AA}$ ($\pm 35$~\% of the lattice constant) imply relatively small energy changes of 150~meV or $\approx 5\,k_\mathrm{B}T$, where $k_\mathrm{B}$ is the Boltzmann constant and  {$T$ the considered temperature.}
Interestingly, we also find that the energy changes due to Br displacements are generally much higher in the static potential well (see Fig.~\ref{fig:invb_atom}), that is when we calculate energy changes of Br displacements for an in-phase octahedral tilting mode in the frozen-phonon picture. 
 {The findings therefore show that including vibrational coupling is concurrent with a lowering of total energy changes that are associated with Br displacements.
This can be explained by the equivalent statement that inlcuding vibrational anharmonicities is concurrent with the system exhibiting larger atomic displacements.}\\
Second, it is very interesting that we  {clearly find temperature-dependent} potential-energy profiles for Br displacements:
 {at $T=425$~K} the dynamic potential well is reminiscent of a double-well potential and qualitatively similar to the static results obtained here and reported in the literature \cite{yang2017,klarbring2019,marronnier2017, marronnier2018,lanigan-atkins2021}; note, however, that the dynamic well is significantly shallower (see inset in Fig.~\ref{fig:invb_atom}).
 {When we increase the temperature to $525$~K, the double-well features disappear and we obtain a potential energy profile that has hardened, which is the canonical behavior of phonon hardening that is expected for a soft mode above the phase-transition temperature.}\cite{dove1993}
 {Fig. S11 in the SIs finds that the 2D histograms of Br displacements change at the increased temperature as well, reflecting the disappearance of the double-well features.}
 {The findings therefore show} that the Br motions occurring along the considered direction, that is perpendicular to the Pb-Br-Pb axis, are related to the soft phonon modes driving the PbBr$_6$ octahedral tilting.\\
\begin{figure}[t]
    \centering
    \includegraphics[width=0.5\linewidth]{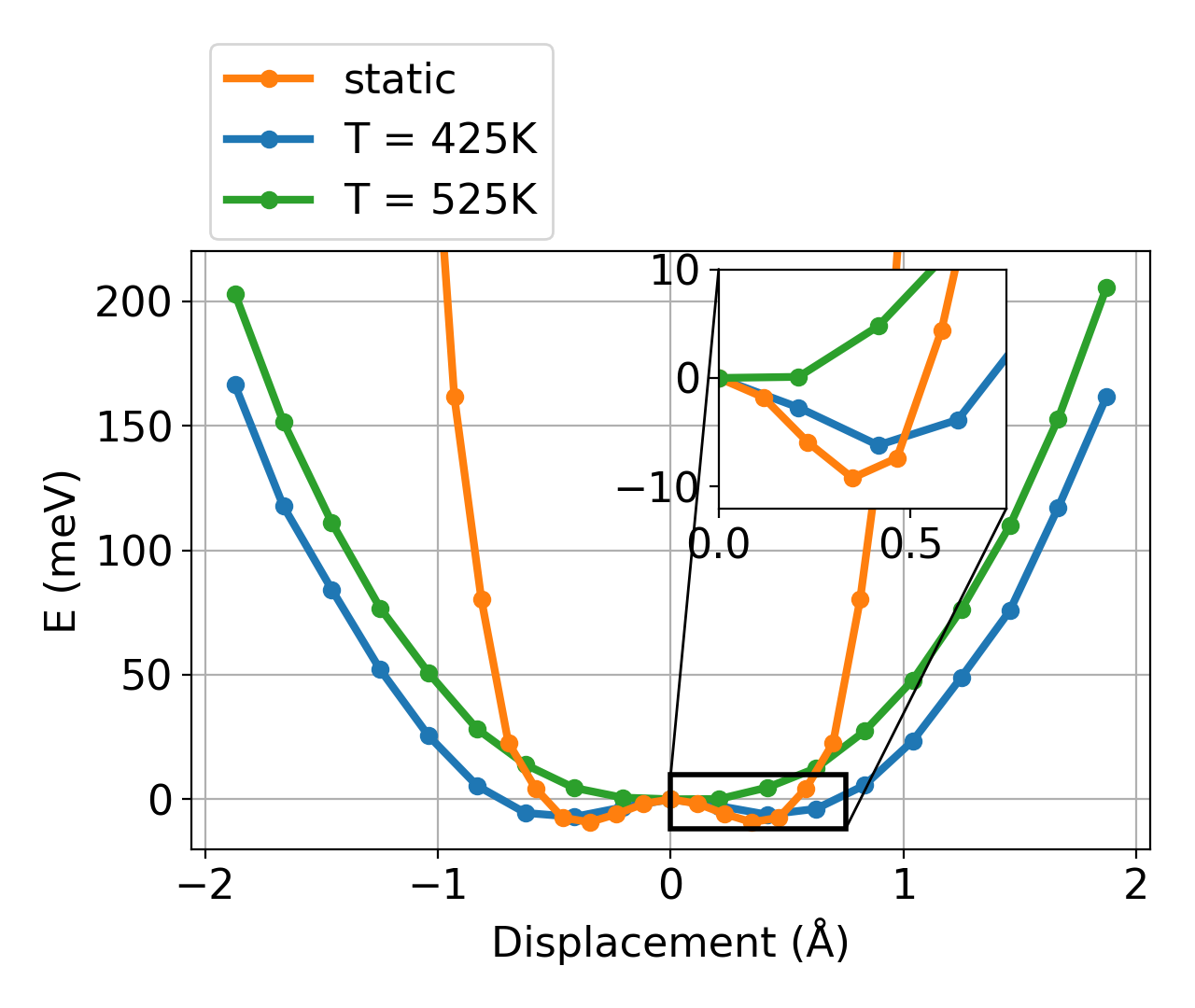}
    \caption{
    Dynamic and static potential well for Br displacements:  {the former were calculated from MD at $T=425$~K (blue curve) as well as $T=525$~K (green curve), and the latter} from a frozen-phonon approach (orange curve), respectively.
    The inset shows a zoom signifying the different depths of  {the potential wells}.
    }
    \label{fig:invb_atom}
\end{figure}
The shallow potential energy profile for the Br displacements and the local Cs-Br coupling motivate further explorations of the energy landscape in CsPbBr$_3$.
We therefore focus on the displacements of a Br atom occurring in the full MD calculation at $425$~K and calculate how the minimum energy in the dynamic potential well changes as a function of the displacement of a nearest-neighbor Cs atom, along the three crystalline directions.
Fig.~\ref{fig:fig6} shows that there is a \textit{multitude of local minima in the Br dynamic potential wells}, depending on displacements of a neighboring Cs atom, in all three directions.
Moreover, these local minima are separated only by few $k_\mathrm{B}T$, which implies that at finite temperature the Br atoms will sample wide ranges of different local configurations over time.
As an example, Cs motion from the most likely position at the center of a nominally cubic cell to one of the 6 side peaks (\textit{cf.} Fig.~\ref{fig:csbr_dist}a) is accompanied by Br motion from one of its many low energy configurations to another (\textit{cf.} Fig.~\ref{fig:csbr_dist}b).
Therefore, for CsPbBr$_3$ the \textit{energy landscape inside the unit cell appears locally and dynamically disordered}, which is in line with the Cs-Br coupling we have found here and the ultrashort correlation lengths in the disorder potential some of us reported previously\cite{gehrmann2019}.\\
\begin{figure}[t]
    \centering
    \includegraphics[width=0.5\linewidth]{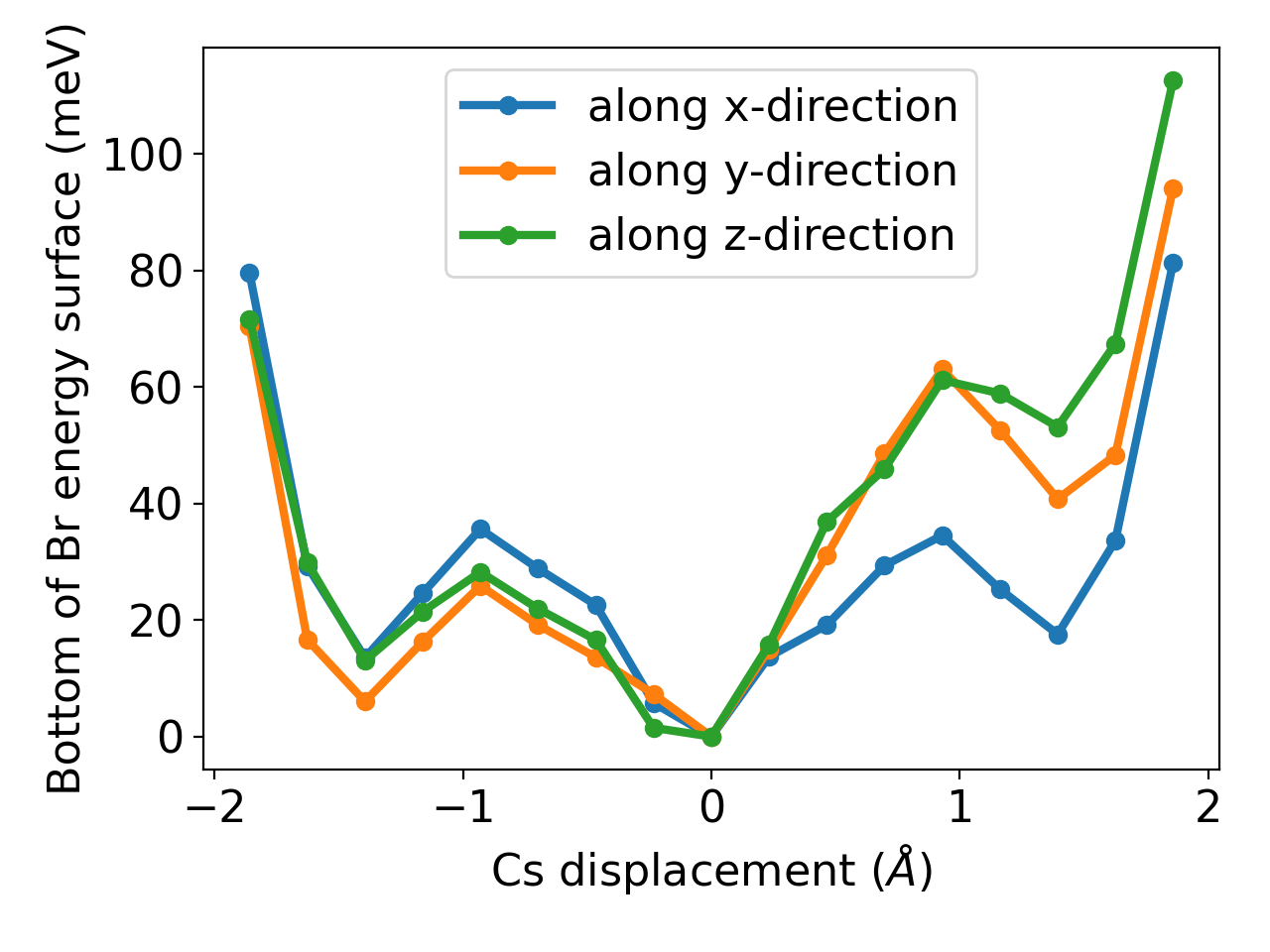}
    \caption{
    Local minimum of the dynamic effective potential, calculated for Br displacement along directions transversal to the Pb-Br-Pb axis, as a function of the Cs displacement. The different curves show the data for Cs displacements occurring along different crystalline directions. Note that the local minima for all potentials wells were calculated up to a common constant (see Methods section for details).
    }
    \label{fig:fig6}
\end{figure}
Furthermore, we found above that the dynamic potential well for the Br displacements  {at $T=425$~K} is qualitatively similar to a static potential profile of a soft phonon mode that drives octahedral tilting, which motivates an investigation of the octahedral dynamics in CsPbBr$_3$ at $425$~K.
Therefore, we project the Br displacements obtained from MD on in-phase and out-of-phase octahedral tilting modes (see Fig.~\ref{fig:inout}a) to find that in-phase tiltings dominate the cubic phase of CsPbBr$_3$~\cite{young2016,klarbring2019}, while out-of-phase tiltings only oscillate around zero.
We probe the role of Cs-Br coupling for octahedral tilting by calculating the cross-correlation of Br displacements involved in octahedral tilting, \textit{i.e.}, the projected displacements shown in Fig.~\ref{fig:inout}a, with specific Cs displacements occurring along the average of the $y$- and the $z$-axis (see Fig.~\ref{fig:inout}b).
From this, significant correlation of 0.6 between Cs displacements and Br displacements associated with in-phase tilting is found, whereas for the out-of-phase tilting displacements it remains close to zero.
These findings demonstrate the important role of Cs-Br coupling in the octahedral tilting dynamics of CsPbBr$_3$. 
Along with related effects, such as the stereochemically active lone-pair that is described in, \textit{e.g.}, Ref. \cite{gao2021}, this motivates a departure from simplified descriptions based purely on geometric factors of the idealized average structure,\cite{goldschmidt1926} as discussed in recent articles.\cite{burger2018,filip2018}
\\
 {We also found above that at $T=525$~K the dynamic potential well for the Br displacements loses the double-well features and phonon hardening occurs. 
Fig. S15 in the SI shows that when we rise the temperature in the MD to $525~\mathrm{K}$, we still find in-phase tilting to be more pronounced than out-of-phase tilting. 
At the same time, at higher temperature the assignment to contributions from specific tilting modes is less clear. 
For example, we show in the SI, Fig. S15, that at $T=525$~K projections of Br displacements onto the in-phase tilting modes that occur around the $z$-direction can, at least for some time of the trajectory, oscillate around zero.
We also show in the Fig. S11 that specific next-nearest-neigbor Br-Br PDFs attain a more Gaussian-like behavior at $T=525$~K compared to $T=425$~K, 
which means that the motions of next-nearest-neigbor Br atoms become less correlated at the higher temperature.
This implies that thermal disorder becomes increasingly more important than correlated tilting motions at higher temperature.
We interpret the findings such that at a temperature significantly above the tetragonal-cubic phase transition of CsPbBr$_3$, the octahedral tilting modes are not fully representative anymore of the local structural fluctuations appearing in the system.
This is indicative of a breakdown of the normal-mode picture that was discussed in a recent Raman spectroscopy study.}\cite{cohen2022}
\\
 {Finally, we mention again that CsPbBr$_3$ in its cubic phase is a prototype case as far as the intriguing structural fluctuations in HaPs are concerned. 
We therefore believe that the effects investigated and discussed in our work, such as the intra unit-cell dynamic disorder and octahedral tilting dynamics, are representative for several variants of the broader class of HaP materials.
Nevertheless, it is important to stress that we do expect ionic substitutions could play an important role too. 
To provide only one but a very prominent example, the substitution of Cs by a polar organic species can induce further stochastic effects to the structural dynamics of the system, which have been extensively studied for a long time.}\cite{poglitsch1987}
 {Such effect can play an important role as well when further structural and ionic variations are considered as is the case in, \textit{e.g.}, double perovskite compounds.}\cite{klarbring2020,cohen2022}

\begin{figure}
    \centering
    \includegraphics[width=0.5\linewidth]{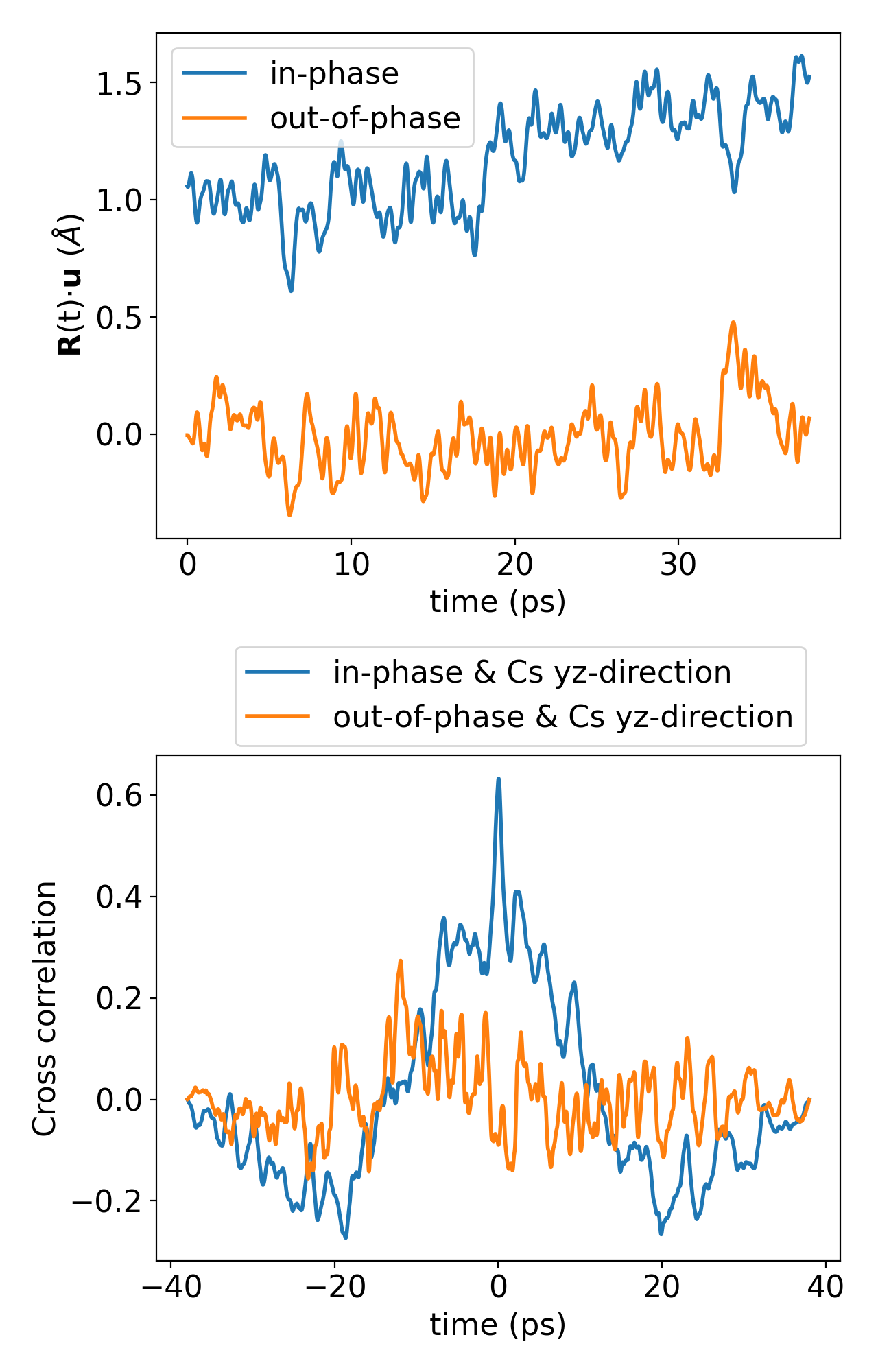}
    \caption{
    \textbf{(a)} Br displacements projected onto the 2 octahedra-tilting modes around $x$-axis, calculated as a function of time in MD at $T=425$~K.  {For projections onto tilting modes occurring around other axes, see SI.}
    \textbf{(b)} Cross-correlation between the Br displacements projected onto the in-phase tilting mode, and the Cs absolute displacement over the plane perpendicular to the Pb-Br-Pb axis.
    Note that the in-phase tilting exhibits a significant correlation with the Cs displacements.
    }
    \label{fig:inout}
\end{figure}

\section*{Conclusions}

In summary, our work focused on the local atomic structure as it occurs in the finite-temperature vibrational dynamics of the HaP material CsPbBr$_3$.
Using first-principles MD performed at $T=425~\mathrm{K}$  {and $T=525~\mathrm{K}$}, we investigated the short-range dynamic interactions occurring among the atoms in the system and found an interlocking of neighboring Cs and Br atoms.
The dynamic coupling of the two species was shown to manifest in the most likely Cs-Br distance being significantly shorter than what is inferred from an ideal cubic structure and to be connected to very shallow potential wells for Br motions, which are found to occur in an energy landscape that is locally and dynamically disordered.
We have probed the disorder in CsPbBr$_3$ to reveal an interesting dynamic coupling of Cs and Br, which we quantified on an atomic scale and showed to be involved in profound structural deviations occurring locally and dynamically in the system.
 {Specifically, we showed that Cs-Br coupling play an important role in the octahedral tilting dynamics of CsPbBr$_3$.}
 {Since the Br atomic orbitals are known to significantly contribute to the electronic states close to the band edges, this is relevant for the optoelectronic properties of HaPs.}
Our findings underline that such dynamic structural effects appear even inside a nominal unit cell of HaPs and, therefore, motivate future work to explore how structural deviations that cannot be inferred from consideration of just the average geometry and small atomic displacements impact optoelectronic and other properties of these materials.

\section*{Theoretical Methods}
DFT-MD calculations were performed with the Vienna Ab-initio Simulation Package (VASP) \cite{kresse1996}, using a $4\times4\times4$ (320 atoms) supercell of cubic CsPbBr$_3$ within the canonical (NVT) ensemble at 425~K.
A Nos\'{e}–Hoover thermostat \cite{nose1984,hoover1985}---as implemented in VASP\cite{kresse1994}---with a 6 fs time step was used to control the temperature.
The DFT calculations performed at each time-step used the projector-augmented wave PAW method \cite{kresse1999} and the code-supplied PAW potentials to describe core-valence electron interactions.
The Perdew-Burke-Ernzerhof (PBE) form of the generalized-gradient approximation for the exchange–correlation functional \cite{perdew1996} was used in conjunction with dispersive-interaction corrections following the Tkatchenko-Scheffler method, \cite{tkatchenko2009} since these are important for capturing accurately structural and dynamical properties of HaPs \cite{egger2014,wang2013,beck2019}.
We employed a plane-wave kinetic-energy cutoff of 300 eV, a single $\mathbf{k}$-point (the $\Gamma$-point), and an energy-convergence threshold of $10^{-6}$ eV.
The MD simulations were equilibrated for 2.5 ps or longer, which was followed by  {production runs of 38 ps ($T=425~\mathrm{K}$) and 39 ps ($T=525~\mathrm{K}$)}. 
The choice of numerical parameters and simulation time was validated by inspection of the power spectrum of the velocity autocorrelation function, using calculations with tighter numerical settings as a reference.\\
Using the data obtained from the MD trajectories, the distribution of Br displacements were calculated as 2D histograms across different planes parallel (see Fig.~\ref{fig:csbr_dist}c) or perpendicular (see Figs.~\ref{fig:csbr_dist}b and \ref{fig:Dyn5d}) to each Pb-Br-Pb bonding axis. Similarly, for the distributions of Cs displacements, 2D histograms have been calculated across the $xy$, $yz$ and $zx$ planes (see Fig.~\ref{fig:csbr_dist}a)
The normalized Cs-Br PDF, $G(d_\mathrm{Cs-Br})$,  {which is shown in} Fig.~\ref{fig:pdf}, was computed using the distance between Cs and 12 nominal nearest-neighbor Br atoms,
averaging over all the Cs atoms in the supercell, $N_\mathrm{Cs}$, and all MD-sampled configurations, $N_t$: 
\begin{equation}
    G(d_\mathrm{Cs-Br}) = \frac{1}{N_\mathrm{Cs}N_{t}}\sum_i \delta (d_\mathrm{Cs-Br}-d_i)
\end{equation}
where $d_i$ is the distance between nearest-neighbor Cs and Br atoms of the $i$-th Cs-Br pair in one MD-sampled configuration.
In Fig. \ref{fig:fig4}, we disentangled the 12 nearest-neighbor Br atoms contributing to this PDF into two sets, which include 8 pairs with larger Cs-Br distance and 4 pairs with a smaller one, and calculated PDFs with the same procedure for these two sets. Note that the atoms included in either set could be different at different MD snapshots.

The distribution of Br displacements transversal to the Pb-Br-Pb axis was calculated in conditional histograms based on 4 scenarios stipulating 4 different types of Cs motions, as summarized in Fig.~\ref{fig:Dyn5d} and described in detail as follows:
\begin{enumerate}[label=(\roman*)]
    \item the first scenario (see Fig.~\ref{fig:Dyn5d}a) considers small Cs displacements in any direction: $$-0.12\mathrm{\AA}\leq x,  y,  z \leq 0.12\mathrm{\AA}.$$
    \item the second scenario (see Fig.~\ref{fig:Dyn5d}b) considers Cs motion along one direction that is parallel to the Pb-Cs-Br axis (\textit{i.e.}, $x$), such that large displacements along that axis and smaller ones along the others are included:
    $$-1.51\mathrm{\AA} \leq  x \leq -1.27\mathrm{\AA}$$ $$-0.12\mathrm{\AA} \leq  y,  z \leq 0.12\mathrm{\AA}$$ or $$1.27\mathrm{\AA} \leq  x \leq 1.51\mathrm{\AA}$$ $$-0.12\mathrm{\AA} \leq  y,  z \leq 0.12\mathrm{\AA}.$$ 
    Only the negative displacements in $ x$ are presented in Fig.~\ref{fig:Dyn5d}b.
    \item the third scenario (see Fig.~\ref{fig:Dyn5d}c) considers Cs motion towards the Pb-Br-Pb axis along one of its perpendicular axes (\textit{i.e.}, $- y$), such that small displacements are considered along the other axes: $$-1.51\mathrm{\AA} \leq  y \leq -1.27\mathrm{\AA}$$ $$-0.12\mathrm{\AA} \leq  x,  z \leq  0.12\mathrm{\AA}.$$
    \item the fourth scenario (Fig.~\ref{fig:Dyn5d}d) considers  Cs motion away from the Pb-Br-Pb axis along one of ity perpendicular axes (\textit{i.e.},$+ y$), such that small displacements are considered along the other axes: $$1.27\mathrm{\AA} \leq  y \leq 1.51\mathrm{\AA}$$  $$-0.12\mathrm{\AA} \leq  x,  z \leq 0.12\mathrm{\AA}.$$
\end{enumerate}
Hence, in order to condition the Br displacement histograms by Cs displacements of a given type, the 3D space was divided into cubes with $\approx0.24~\mathrm{\AA}$ length to record the Cs positions.
For these calculations, the data set had to be increased in order to improve the statistics for each motion type. 
Therefore, in addition to the 38 ps trajectory  {at $425$ K} described above we calculated further independent trajectories, namely another two longer one ($\approx$ 30 ps) and four shorter ones ($\approx$ 10 ps each).
Furthermore, in this procedure we have mapped the data of symmetry-equivalent Cs-Br pairs into the statistics of one Cs-Br pair as described in Fig. S4 of the SI.\\
The static, frozen-phonon potential well was calculated using a $2\times2\times2$ cubic supercell, a $\Gamma$-point centered $3\times3\times3$ $\mathbf{k}$-point grid, and by considering displacements of the lowest-energy phonon mode at the $M$-point of the cubic Brillouin zone.
The dynamic potential well was calculated using inversion of a histogram of displacements, $n(\delta r_1, \delta r_2, \delta r_3)$. 
According to Boltzmann statistics, $n(\delta r_1, \delta r_2, \delta r_3) \propto \text{exp}\bigl\{-E(\delta r_1, \delta r_2, \delta r_3)/k_B T\bigr\}$, where $\delta r_1$ is the displacement along $x$-direction (parallel to the bonding axis) and $\delta r_2$, $\delta r_3$ are displacements along $y$ and $z$-directions (perpendicular to the bonding axis). 
Consequently, the potential can be obtained as: $E(\delta r_1, \delta r_2, \delta r_3) \propto - \text{ln}\bigl\{n(\delta r_1, \delta r_2, \delta r_3) \bigr\} \cdot k_B T$. 
The effective potential energy of Br displacement perpendicular to the Br-Pb-Br bonding axis  was calculated by taking the average over two directions: $E_{trans}(r) = (E(0,r,0)+E(0,0,r))/2$. 
The same method was used to obtain the effective energy landscape of one Cs-Br pair in Fig. \ref{fig:fig6}, but based on the 5D histogram recording Cs displacements in 3D Cartesian directions and the transversal displacements of 2 Br atoms. 
For each point in the figure, one Cs displacement along $x$/$y$/$z$ direction was chosen from that distribution, and the lowest energy was taken from the most likely Br transversal displacement at that specific Cs displacement.
It is noted that the dynamic potential well must be viewed as an effective potential for one Cs-Br pair, while the entirety of the energy landscape is significantly more complex since there are other atoms around each Br that in general will interfere with its energetic surroundings in a highly non-trivial manner.\\
The quantity $\mathbf{R}(t)\cdot \mathbf{u}$ was calculated by projecting the Br displacement vector, $\mathbf{R}(t)$, onto the in-phase and out-of-phase octahedral tilting mode $\mathbf{u}$, which are shown in Fig. S13 of the SI.
The $x$-direction was set to be the rotation axis and $\mathbf{R}(t)$ contains the displacement of all Br atoms in the supercell at time $t$. 
The normalized cross-correlation was calculated between the absolute value of in-phase/out-of-phase projections, $|\mathbf{R}(t)\cdot \mathbf{u}|=A(t)$, and the average of the absolute Cs displacement along $y$- and $z$-directions, $B(t)$. The normalized cross-correlation between the two variables $A(t)$ and $B(t)$ was then calculated as:
\begin{equation}
    C(\Delta t) = \frac{1}{N_t}\sum_{t=0}^{t=N_t}\frac{(A(t+\Delta t)-\Bar{A})\cdot (B(t)-\Bar{B})}{\sigma(A)\cdot\sigma(B)}
\end{equation}
where $\Bar{A}/\Bar{B}$, $\sigma(A)/\sigma(B)$ are the mean values and the standard deviations of $A$/$B$.

\section*{Acknowledgements}
The authors thank Olle Hellman and Omer Yaffe for fruitful discussions.
Funding provided by the Alexander von Humboldt-Foundation in the framework of the Sofja Kovalevskaja Award, endowed by the German Federal Ministry of Education and Research, by
the Deutsche Forschungsgemeinschaft (DFG, German Research Foundation) via SPP2196 Priority Program (project-ID: 424709454) and via Germany's Excellence Strategy - EXC 2089/1-390776260, and by the Technical University of Munich - Institute for Advanced Study, funded by the German Excellence Initiative and the European Union Seventh Framework Programme under Grant Agreement No. 291763, are gratefully acknowledged.
The authors further acknowledge the Gauss Centre for Supercomputing e.V. for funding this project by providing computing time through the John von Neumann Institute for Computing on the GCS Supercomputer JUWELS at J\"ulich Supercomputing Centre.

\section*{Supporting Information}
 {Additional analysis of atomic distribution functions; pair distribution functions for different atomic pairs at $425$~K and $525$~K; conditional 2D Br histograms prior to symmetry mapping; analysis of relation between Br displacements and octhedral rotations; additional analysis of contribution from Br atoms to Cs-Br pair distribution function; projections onto in-phase and out-of-phase mode around three axis at $425$~K and $525$~K}

\printbibliography
\end{document}